\documentclass[11pt,a4paper]{article}
\usepackage{jheppub}
\usepackage[T1]{fontenc}

\notoc    

\title{Addendum: Observables for General Relativity related to geometry}

\author[a]{Pawe{\l} Duch,}
\author[b]{Wojciech Kami\'nski,}
\author[b]{Jerzy Lewandowski}
\author[b]{and J\k{e}drzej \'Swie\.zewski}

\affiliation[a]{Institute of Physics, Jagiellonian University,\\ {\L}ojasiewicza 11, 30-348  Krak\'{o}w, Poland} 
\affiliation[b]{Faculty of Physics, University of Warsaw,\\ Pasteura 5, 02-093 Warszawa, Poland}

\emailAdd{pawel.duch@uj.edu.pl} 
\emailAdd{wojciech.kaminski@fuw.edu.pl} 
\emailAdd{jerzy.lewandowski@fuw.edu.pl}
\emailAdd{swiezew@fuw.edu.pl}

\abstract{
In this addendum we clarify a point which strengthens one of the results from \cite{JHEP}. Namely, we show that the algebra of the observables $F(r,\theta)$ is yet simpler then it was described in \cite{JHEP}. This is an important point, because with this simplification an important subalgebra becomes canonical, allowing for a natural reduction of the phase space.
}
      
\begin{document}
\maketitle

\section{Introduction}

In equations (4.11) from \cite{JHEP} we computed the Poisson brackets between observables not constructed from the momentum conjugate to the metric (i.e., $\Phi_\alpha, \Pi^\alpha, Q_{AB}$) with the observables corresponding to that momentum (i.e., $P^{AB}$).\footnote{Note that we are adopting the notation from \cite{JHEP} throughout this addendum. This means in particular, that the indeces $A,B,\ldots$ correspond to the angular components of the appriopriate tensors.} In each case the result contained terms proportional to a delta at zero. In fact, this result can be strengthened a bit, by realizing that in an important class of cases the terms proportional to the delta at zero on the right-hand sides of (4.11) do not contribute and therefore are zero. This is an important simplification of the algebra, because without those delta terms and knowing additionally that the Poisson bracket $\{P^{AB},\ P^{CD}\}$ is vanishing the algebra of the observables $\Phi_\alpha, \Pi^\alpha, Q_{AB}, P^{AB}$ is truly canonical.  

In this addendum, we will first analize the expressions in (4.11) and show that the vanishing of those terms can be seen already at this level. In fact, a closer look at (4.11a) will be enough to provide an argument for the rest of the equations. Next, we will compute the Poisson bracket $\{P^{AB},\ P^{CD}\}$.

\section{Vanishing of the $\delta(r')$ terms}

Let us recall the result spelled out in (4.11a)
\begin{equation}\label{eqfromJHEP}
\{ \Phi_\alpha(r,\theta),\ P^{AB}(r',\theta')\}\ =\ -\frac{1}{2} y'^A_{,I} y'^B_{,J} T^{IJ}_{KL} h^{LM}x^K y^C_{,M} \partial_C\Phi_\alpha(r,\theta)\delta(r').
\end{equation}
Notice, that this equation, when smeared with a smooth tensor field $w_{IJ}$, such that 
\begin{equation}\label{wrazero}
w_{ra} = 0
\end{equation}
gives
\begin{equation}\label{eq2}
\int dr'd^2\theta' w_{AB}(r',\theta')\{ \Phi_\alpha(r,\theta),\ P^{AB}(r',\theta')\}\ =\ -\frac{1}{2} w_{IJ}(0) T^{IJ}_{KL} h^{LM}x^K y^C_{,M} \partial_C\Phi_\alpha(r,\theta).
\end{equation}
However, for fields $w_{IJ}$ satisfying condition \eqref{wrazero} we have
\begin{equation}
w_{IJ}(0)\ =\ \underset{r\rightarrow0}{\lim}\Big(n_I(\theta) n_J(\theta) w_{rr}(r,\theta)\Big)\ =\ 0,
\end{equation}
and therefore the right hand side of \eqref{eq2} vanishes. Since the components $w_{AB}$ were arbitrary we conclude that in fact
\begin{equation}
\{ \Phi_\alpha(r,\theta),\ P^{AB}(r',\theta')\}\ =\ 0.
\end{equation}

This fact can also be phrased in the following way. The left hand side of \eqref{eqfromJHEP} is smeared in primed variables with tensor fields vanishing at zero (since for any smooth tensor field $w_{IJ}$ we have $w_{AB}\sim r^2$, for a smooth tensor field satisfying additionally equation \eqref{wrazero} we have $w_{AB}\sim r^3$) and therefore the delta at zero does not contribute to the result. One might worry that when we consider observables which contain radial derivatives of the momentum, namely $\partial_r^{(n)}P^{AB}$, we will see some contributions. It turns out, that for smooth tensor fields $w_{IJ}$ satisfying \eqref{wrazero} which are symmetric (namely $w_{IJ}=w_{JI}$, which is the case here since $P^{IJ}$ is symmetric) we have $w_{AB}\sim r^4$, therefore even for observables containing the derivative $\partial_r P^{AB}$ the delta term does not contribute in their Poisson brackets.

The behaviour at zero of the $w_{IJ}$ used above can be justified in the following way. For a smooth tensor field $w_{IJ}$ satisfying \eqref{wrazero}, we can infer 
\begin{equation}
	w_{IJ}(0)=0.
\end{equation}
This implies 
\begin{equation}
	w_{IJ}=x^Kf_{KIJ},
\end{equation}
for some smooth tensor field $f_{KIJ}$. On one hand, from \eqref{wrazero} we get 
\begin{equation}
	x^Kx^If_{KIJ}=0,
\end{equation}
and therefore, taking the second derivative at zero, we get 
\begin{equation}
	f_{(KI)J}(0)=0,
\end{equation}
which means
\begin{equation}\label{propertyOne}
	f_{KIJ}(0)=f_{[KI]J}(0).
\end{equation}
On the other hand,
\begin{equation}
	f_{KIJ}(0)=\partial_Kw_{IJ}(0)
\end{equation}
and hence, invoking the fact that $w_{IJ}$ is symmetric,
\begin{equation}\label{propertyTwo}
	f_{KIJ}(0)=f_{K(IJ)}(0).
\end{equation}
Combining \eqref{propertyOne} and \eqref{propertyTwo} we get that
\begin{equation}
	f_{KIJ}(0)=0,
\end{equation}
which means
\begin{equation}
	\partial_Kw_{IJ}(0)=0
\end{equation}
justifying the statement that for the tensor fields $w_{IJ}$ of interest here, we have $w_{AB}\sim r^4$.

It is straightforward to apply the above arguments in all of the equations from (4.11) leaving them in the form
\begin{subequations}
\begin{align}
\{ \Phi_\alpha(r,\theta),\ P^{AB}(r',\theta')\}\ =&\ 0,\\
\{ \Pi^\alpha(r,\theta),\ P^{AB}(r',\theta')\}\ =&\ 0,\\
\{Q_{CD}(r,\theta),\ P^{AB}(r',\theta')\}\ =&\ \delta^A_{(C} \delta^B_{D)} \delta(r - r')\delta(\theta - \theta').
\end{align}
\end{subequations}

\section{The Poisson bracket $\{P^{AB},\ P^{CD}\}$}

Let us compute the following Poisson bracket
\begin{align}
\{P^{AB}(r',\theta'),\ \int d^3x w_{IJ}(x)P^{IJ}(x)\}\ &= \ \int d^3\sigma \int d^3x \frac{\delta D_{P^{AB}(r',\theta')}}{\delta q_{ra}(\sigma)}\frac{\delta P^{IJ}(x)}{\delta p^{ra}(\sigma)}w_{IJ}(x)\\
&=\ \int d^3\sigma \frac{\delta D_{P^{AB}(r',\theta')}}{\delta q_{ra}(\sigma)}w_{ra}(\sigma).
\end{align}
Choosing $w_{IJ}$ to be such a tensor field that the $w_{ra}$ components vanish, we obtain
\begin{equation}
\{P^{AB}(r',\theta'),\ P^{CD}(r,\theta)\}\ =\ 0.
\end{equation}
Recalling equations (4.5) from \cite{JHEP}, namely,
\begin{subequations}
\begin{align}
\{ \Phi_\alpha(r,\theta),\ \Pi^{\alpha'}(r',\theta')\}\ &=\ \delta_\alpha^{\alpha'}\delta(r - r')\delta(\theta - \theta'),\\
\{\Phi_\alpha(r,\theta),\ \Phi_{\alpha'}(r',\theta')\}\ &=\ \{\Pi^\alpha(r,\theta),\ \Pi^{\alpha'}(r',\theta')\}\ =\ \{Q_{AB}(r,\theta),\ Q_{CD}(r',\theta')\}\ =\ 0,\\
\{Q_{AB}(r,\theta),\ \Phi_\alpha(r',\theta')\}\ &=\ \{Q_{AB}(r,\theta),\ \Pi^\alpha(r',\theta')\}\ =\ 0,
\end{align}
\end{subequations}
we conclude that the algebra of observables $\Phi_\alpha, \Pi^\alpha, Q_{AB}, P^{AB}$ is canonical.

\section{Final remarks}

To see the consequences of the simplification of the Poisson algebra presented above, we should adress its applications. Note, that in the ADM formulation of General Relativity we are interested in, the vector constraint can be written in the form
\begin{equation}
C[\vec N]\ =\ 2\int P^{IJ} \nabla_I N_J,
\end{equation}
so the field smearing the momentum is in this case $2\nabla_{(I}N_{J)}$.  Though in general the value of $2\nabla_{(I}N_{J)}$ at zero is dictated by the shift vector and may be nonvanishing, it turns out, that a gauge fixing of General Relativity related to the observables discussed here is possible in which the shift vector is fixed in such a way that exactly $2\nabla_{(I}N_{J)}(0) = 0$. Following the discussion presented above we see that the condition $w_{IJ}(0)=0$ is enough to draw the conclusion that the right hand side of \eqref{eqfromJHEP} does not contribute to the Poisson bracket of $\Phi_\alpha$ with the vector constraint. Arguing in the same manner  we can see that the delta term does not contribute in the Poisson brackets of the vector constraint with all the observables.

The strength of the observables constructed in \cite{JHEP} is that they have a very clear geometrical interpretation. On the other hand, they depend nonlocally on the canonical data and this might be worrisome when one thinks of their possible applications. However, a reduction of the phase space to observables $\Phi_\alpha, \Pi^\alpha, Q_{AB}, P^{AB}$ advocated in \cite{JHEP} and mentioned in the previous paragraph can be completed. As this addendum shows, the reduced variables can have canonical Poisson brackets, and the nonlocality will be reappearing only in the Hamiltonian. The detailed construction of the reduced phase space is presented in \cite{RGI}. Moreover, the reduced phase space can be used in quantisation, which is discussed in \cite{RGII} and \cite{RGletter}.

\section{Acknowledgements}

PD acknowledges the support in a form of a scholarship of the Marian Smoluchowski Krak\'{o}w Scientific Consortium Matter-Energy-Future from KNOW funding. This work was partially supported by the grant of Polish National Science Centre nr 2011/02/A/ST2/00300 and by the grant of Polish National Science Centre nr 2013/09/N/ST2/04299. J\'S would like to thank Norbert Bodendorfer for helpful discussions.

\end{document}